\documentclass[acmsmall]{acmart}

\AtBeginDocument{%
  }

\setcopyright{acmlicensed}
\acmJournal{PACMCGIT}
\acmYear{2025} \acmVolume{8} \acmNumber{1} \acmArticle{10} \acmMonth{5}\acmDOI{10.1145/3728299}
\usepackage{acmart-taps}

\begin{document}

\title{Aokana: A GPU-Driven Voxel Rendering Framework for Open World Games}

\author{Yingrong Fang}
\affiliation{%
  \institution{Fudan University}
  \department{Shanghai Key Laboratory of Data Science, School of Computer Science}
  \city{Shanghai}
  \country{China}
  }
\email{23210240147@m.fudan.edu.cn}

\author{Qitong Wang}
\authornote{Qitong Wang is the corresponding author.}
\affiliation{%
  \institution{Harvard University}
  \city{Cambridge}
  \country{USA}}
\email{qitong@seas.harvard.edu}

\author{Wei Wang}
\affiliation{%
  \institution{Fudan University}
  \department{Shanghai Key Laboratory of Data Science, School of Computer Science}
  \city{Shanghai}
  \country{China}
  }
\email{weiwang1@fudan.edu.cn}

\begin{abstract}
Voxels are among the most popular 3D geometric representations today. 
Due to their intuitiveness and ease-of-editing, voxels have been widely adopted in stylized games and low-cost independent games. 
However, the high storage cost of voxels, along with the significant time overhead associated with large-scale voxel rendering, limits the further development of open-world voxel games. 
In this paper, we introduce \textbf{Aokana}, \textbf{a} GPU-Driven V\textbf{o}xel Rendering Framewor\textbf{k} for
Ope\textbf{n} World G\textbf{a}mes.
Aokana is based on a Sparse Voxel Directed Acyclic Graph (SVDAG).
It incorporates a Level-of-Details (LOD) mechanism and a streaming system, enabling seamless map loading as players traverse the open-world game environment.
We also designed a corresponding high-performance GPU-driven voxel rendering pipeline to support real-time rendering of the voxel scenes that contain tens of billions of voxels. 
Aokana can be directly applied to existing game engines and easily integrated with mesh-based rendering methods, demonstrating its practical applicability in game development.
Experimental evaluations show that, with increasing voxel scene resolution, Aokana can reduce memory usage by up to ninefold and achieves rendering speeds up to 4.8 times faster than those of previous state-of-the-art approaches.
\end{abstract}




\begin{teaserfigure}
  \centering
  \includegraphics[width=\linewidth]{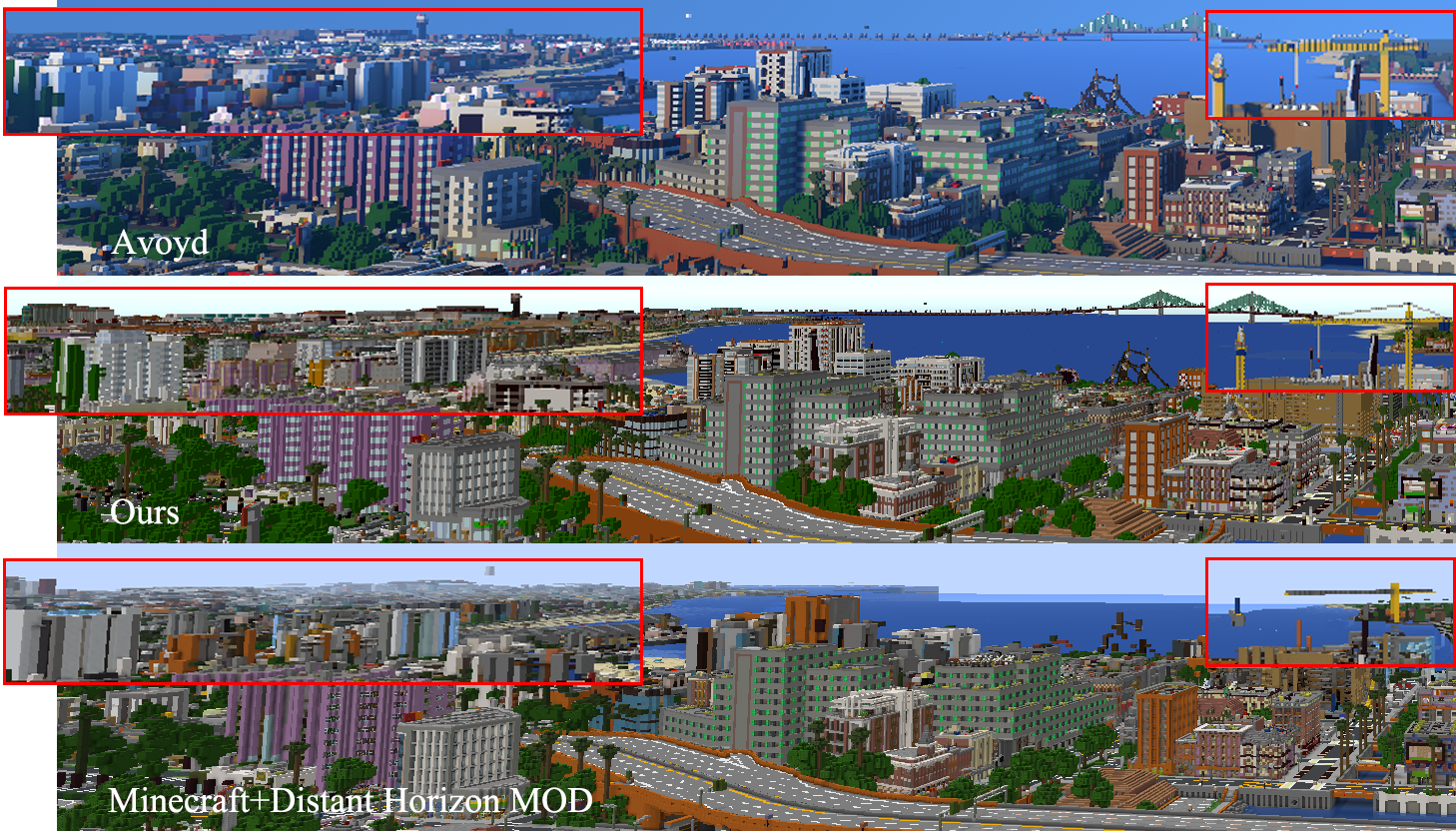}
  \caption{
  A visual comparison of the level of details between our method and the existing gaming softwares. 
  Note that the difference of the sky, as a result of different shading implementations, is irrelevant of the voxel rendering algorithms.
  Our method can preserve the details of distant objects better than Avoyd and Minecraft+Distant Horizon MOD.
  }
  \label{fig:lod_cmp}
  \Description{LOD Comparison}
\end{teaserfigure}

\maketitle

\section{Introduction}
A voxel serves as the three-dimensional equivalent of a pixel, representing a value within a regular grid in three-dimensional space.  
As one of the key representations of three-dimensional data, voxels are widely utilized across various fields, including data visualization, industrial software, and video games.

In recent years, open-world games have become increasingly popular among players.
However, the large-scale maps typically required for open-world games necessitate substantial costs and extensive human resources for asset production, which deters many low-budget teams and independent game developers from pursuing this genre. 
The voxel representation of 3D geometry is highly intuitive and structured, facilitating the bulk generation of voxel assets through randomization or procedural methods. This capability enables developers to easily construct vast maps. 
As a result, more independent developers are turning their attention to stylized voxel games, such as Minecraft and Tear Down.
Taking Minecraft as an example, where each block is sized at $1m^3$, an almost infinite natural environment map is generated using random noise, effectively circumventing a significant portion of the storage costs associated with the map.
However, this approach is not always feasible. 
For instance, when developers aim to create a large artificial city map, they need to pre-store the complete map data on disk and carefully select the optimal moment to load the necessary data into both memory and VRAM.
If the game requires finer voxel details (e.g., individual voxel sizes smaller than $1m^3$), the storage costs will further escalate. 
The challenge of efficiently storing and rendering vast amounts of voxel data in real-time on consumer-grade hardware constrains the further development of voxel-based games within the open-world genre.

Due to the limitations imposed by the size of VRAM on consumer GPUs, it is typically not feasible to load the raw data of an entire open-world game map into GPU memory. 
In gaming, there are primarily two approaches to address this issue. 
The first approach involves reducing the player's viewing distance; distant areas are obscured by fog, allowing only the nearby visible region to be loaded instead of the entire map. 
The second approach entails decreasing the detail of distant objects or culling smaller objects at a distance, often utilizing Level of Detail (LOD) systems such as Hierarchical Level of Detail (HLOD), which can reduce the memory footprint for distant objects. 
Modern games commonly employ a combination of these two strategies. In open-world games, to allow players to appreciate distant scenery, it is usually necessary to provide a large viewing distance. 
Our goal is to ensure that the entire open-world game map is always loaded while maintaining good detail for distant objects and supporting efficient real-time rendering.

In previous research, the compression of geometric and color information for voxels has been extensively studied. 
Building on this foundation, our work shifts the focus towards enhancing the rendering performance and VRAM usage of large-scale voxel scenes. 
Many earlier methods either required storing the complete scene data in VRAM, leading to high overhead, or utilized deep data structures connected by pointers, which are not cache-friendly. 
The use of these deep data structures results in decreased VRAM access performance when the voxel scene resolution is high, ultimately causing a loss of rendering performance.

To address these issues, we developed a streaming loading system that does not require loading the entire voxel scene data into VRAM. 
We utilize multiple shallow SVDAGs to maintain the entire scene, and by employing Hi-Z occlusion culling and a visibility buffer, we minimize memory access frequency and overdraw.

For voxel scenes with a resolution above 32K, our method achieves a rendering speed 2 to 4 times faster than HashDAG, and only about $5\%$ of the complete voxel scene data needs to be loaded into VRAM when navigating the scene.

Our main contribution is the introduction of a GPU-Driven voxel rendering framework for open-world games, that can enable efficient real-time rendering of large-scale voxel environments on consumer-grade hardware, incorporating voxel LOD and streaming system. 
Additionally, since our voxel rendering pipeline simply adds several passes to the forward rendering pipeline, our approach is compatible with existing mesh-based rendering pipelines.
We have implemented our approach in the popular commercial game engine Unity to demonstrate its practical applicability in real-world game projects.

\section{Related Work}
\subsection{Rasterization-based Methods}

Viewing voxels as triangle meshes is the most classical approach and has been widely adopted in contemporary video games. For example, in the renowned video game Minecraft, the world is divided into chunks measuring $16\times 16\times (256\sim 384)$. 
Invisible voxel surfaces are culled, and a series of identical material flat voxel surfaces are merged to reduce the number of vertices and triangles within each chunk. 
In the original version of Minecraft, there is no LOD system; instead, chunks near the player are dynamically loaded based on a predefined viewing distance, while chunks beyond this distance are unloaded. 
Recently, a group of gaming enthusiasts developed the Distant Horizons MOD~\cite{DistantHorizons}, which implements an LOD system for Minecraft by aggregating voxels and then regenerating meshes to construct chunk LODs. 
However, since meshes require vertex, index, and UV information, the storage costs are relatively high, and the large number of triangles leads to increased rasterization overhead, limiting the ability to further enhance LOD accuracy.

Majercik et al.~\cite{Majercik2018Voxel} proposed an efficient Ray-Box Intersection Algorithm that supports real-time rendering of large quantities of dynamic voxels. 
They first splat voxel billboards to provide a rough visibility estimate, and then perform ray-box intersection in the fragment shader to compute precise visibility. 
Their approach does not utilize precomputation techniques or spatial data structures, making it suitable for dynamic scenes where each voxel may change every frame. 
However, in open-world voxel scenarios, static voxels constitute the vast majority, rendering their method less applicable to our context. Additionally, they did not compress voxel data, which limits the scale of voxels that can be stored in VRAM.

Epic Games developed the Nanite Virtual Geometry System~\cite{karis2021nanite} in Unreal Engine 5, which divides meshes into clusters and implements seamless LOD transitions at the cluster level. 
It employs compute shaders for soft rasterization of micro-triangles, creating an efficient GPU-driven rendering pipeline that supports real-time rendering of vast numbers of triangles. 
However, if a scene contains many discontinuous voxels, Nanite may generate an excessive number of disconnected clusters that are hard to be merged.
These disconnected clusters also have poor geometric quality because the construction of LOD relies on the mesh simplification algorithm.
Whether Nanite should be used to construct voxel games remains to be further tested.

Virtual Voxel~\cite{10.1145/3588028.3603664} uses techniques similar to Virtual Texture~\cite{10.1145/2343483.2343488} to select the appropriate LOD for voxel chunks based on the camera's position and to perform streaming loading. 
They treat voxels as a point cloud, initially rendering points containing voxel depth and position information into a 64-bit Visibility Buffer. 
Subsequently, they employ a pass to retrieve voxel position information from the Visibility Buffer, using a compute shader to reconstruct twelve triangles for each individual voxel through soft rasterization. However, they directly store voxel data linearly, resulting in high storage costs, and when the camera approaches the voxel, the number of triangles on a single voxel can far exceed the pixel size, causing a significant decrease in the performance of soft rasterization and leading to performance losses.

\subsection{Ray Marching-based Methods}

Laine et al.~\cite{laine2010efficient} proposed the Efficient Sparse Voxel Octree (ESVO), which stores voxel scenes using a Sparse Voxel Octree structure to reduce voxel storage costs. 
Leveraging this acceleration structure, they designed a ray casting algorithm that can efficiently run on the GPU. They also introduced beam optimization, where a lower-resolution distance image is rendered first, and then, when rendering the full-resolution image, the starting positions of the rays are adjusted based on this distance image. 
This allows the rays to skip most of the empty space in the scene and commence ray marching from surfaces close to the voxels, effectively enhancing the efficiency of the ray marching process.

Kampe et al.~\cite{kampe2013high} proposed the Sparse Voxel Directed Acyclic Graph (SVDAG), which merges isomorphic subtrees of the Sparse Voxel Octree to form a directed acyclic graph (DAG). 
Although this method requires explicit storage of child pointers for each node, the significant merging of many subtrees, particularly leaf nodes, allows for a higher compression ratio compared to ESVO. 
The spatial location information of the original SVO nodes is now implicitly contained in the paths originating from the source point in the SVDAG, enabling efficient ray marching without the need for decompression. 
However, they only compressed geometric information and did not address the handling and compression of color data.

Dolonius et al.~\cite{dolonius2017compressing} successfully decoupled geometric and color data in the SVDAG by additionally recording the number of nodes in each subtree at every node and storing the color of each node in a color array in depth-first search (DFS) order, allowing the SVDAG to support color. 
They also discussed two strategies for compressing color information. One strategy involves mapping the 1D color array to a 2D image using space-filling curves, followed by the application of traditional image compression algorithms, which facilitates offline storage and network transmission. 
The other strategy introduces an additional low-bitrate weight array, while segmenting the color array into several attribute blocks to form an attribute block array. 
Each attribute block stores only two colors, $c_0$ and $c_1$, with all colors in the same block interpolated from $c_0$ and $c_1$ based on the weight. 
During ray casting, a binary search is performed on the attribute block array according to the DFS order of the hit voxel to locate the corresponding attribute block, from which the color of the corresponding voxel can be obtained through interpolation. 
However, they did not further discuss the performance of ray marching.

Richermoz et al.~\cite{richermoz2024gigavoxels} proposed GigaVoxels DP, a volume rendering method that leverages GPU dynamic parallelism. 
They treat $8^3$ voxels as a single brick and store all currently visible bricks in a brick pool, which is a large 3D texture. A ray payload buffer, sized to match the screen, records the current state of each ray at every pixel, including the distance traveled by the ray, the currently accumulated RGBA color, and a boolean lock to prevent concurrent execution on the same pixel. 
When a ray at a pixel determines that it has not hit the expected brick, it triggers a Brick Production thread to load the corresponding brick into the Brick Pool. 
This method uses bricks to store voxels without further compression, resulting in substantial memory overhead. Moreover, the approach assumes that all voxels are transparent, which does not align with the fact that most objects in open-world games are opaque. 
For the scenarios we are studying, this could result in missed optimization opportunities.

V. Careil et al.~\cite{careil2020interactively} built upon the SVDAG framework to introduce persistent functionality with HashDAG, enabling users to perform interactive modifications on large-scale voxel scenes. 
Each modification operation adds a new root node representing the current version of the voxel data, while maximizing the reuse of information from historical versions to reconstruct the tree chain originating from this root node. 
They also designed a hashing mechanism for efficient node queries and a virtual memory system that utilizes fixed, predefined virtual memory addresses, allowing for updates to the HashDAG without the need to move large amounts of data. 
Their implementation similarly stores all data in GPU memory, resulting in high memory requirements. Their method faces the same challenges encountered by all Octree or DAG-based approaches when dealing with large voxel data scales. 
Specifically, as the depth of the data structure increases, the number of indirect jumps during queries also rises, which reduces cache hit rates and leads to performance degradation.

OpenVDB~\cite{museth2019openvdb} is an open source C++ library that implements a novel hierarchical data structure with a large suite of tools for the efficient storage and manipulation of sparse volumetric data discretized on three-dimensional grids. 
This technology is currently widely used in the simulation and rendering of effects such as water, fire, smoke, and clouds, which rely on sparse volumetric data. 
However, further design is still needed to modify the LOD and streaming strategies to meet the demands of voxel-based open-world games.

\section{Method}
We now describe the workflow of our framework. In the preprocessing stage, we divide the world into several chunks and utilize a Sparse Voxel DAG to compress these chunks (see Section 3.1-3.3). 

The real-time rendering stage is depicted in Figure~\ref{fig:pipeline}. We insert the voxel rendering operation between the opaque pass and transparent pass of the forward rendering pipeline (see Section 3.5). 
First, the world is segmented into multiple chunks, and in the chunk selection pass, we filter out the chunks that may be drawn. 
The screen is then divided into $8\times 8$ tiles. 
Second, in the tile selection pass, we utilize the previous frame's Hi-Z texture to perform Hi-Z occlusion culling, generating a list of chunks that may contribute to each tile, resulting in a series of tile-chunk pairs. 
Third, during the ray marching pass, we perform ray marching on the SVDAG to find the intersection points of rays projected from each pixel on the tile and the chunks, writing the intersection information into a 64-bit visibility buffer. 
We then construct the Hi-Z texture for the current frame and re-execute the tile selection pass to identify tile-chunk pairs that were erroneously culled, followed by another round of ray marching. 
Finally, in the Color Resolve Pass, we compute the color and depth information from the 64-bit visibility buffer, writing them to the color target and depth target, respectively. 

We have implemented a Level of Detail (LOD) mechanism and a streaming system (see Section 3.4) to ensure that only the necessary chunks are loaded into VRAM, thereby minimizing VRAM usage.

\aptLtoX[graphic=no,type=html]{\begin{figure}
  \centering
    \includegraphics[width=.5\textwidth]{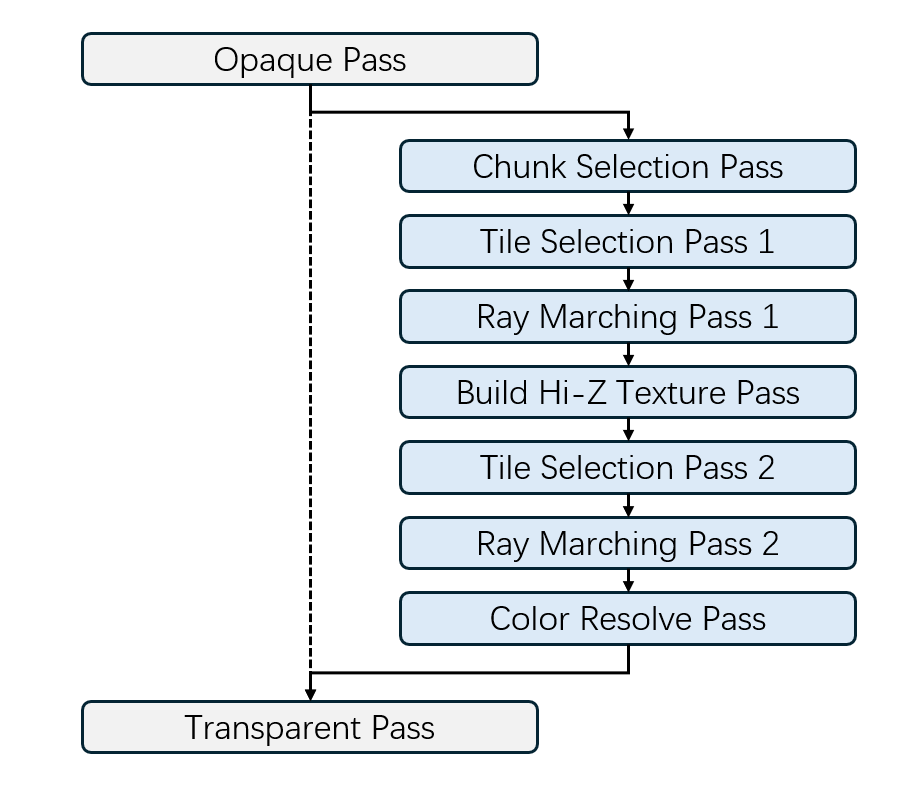}
    \caption{The rendering pipeline overview. Our voxel rendering passes are inserted between the opaque pass and transparent pass of the forward rendering pipeline.}
    \label{fig:pipeline}
\end{figure}
\begin{figure}
    \includegraphics[width=.5\textwidth]{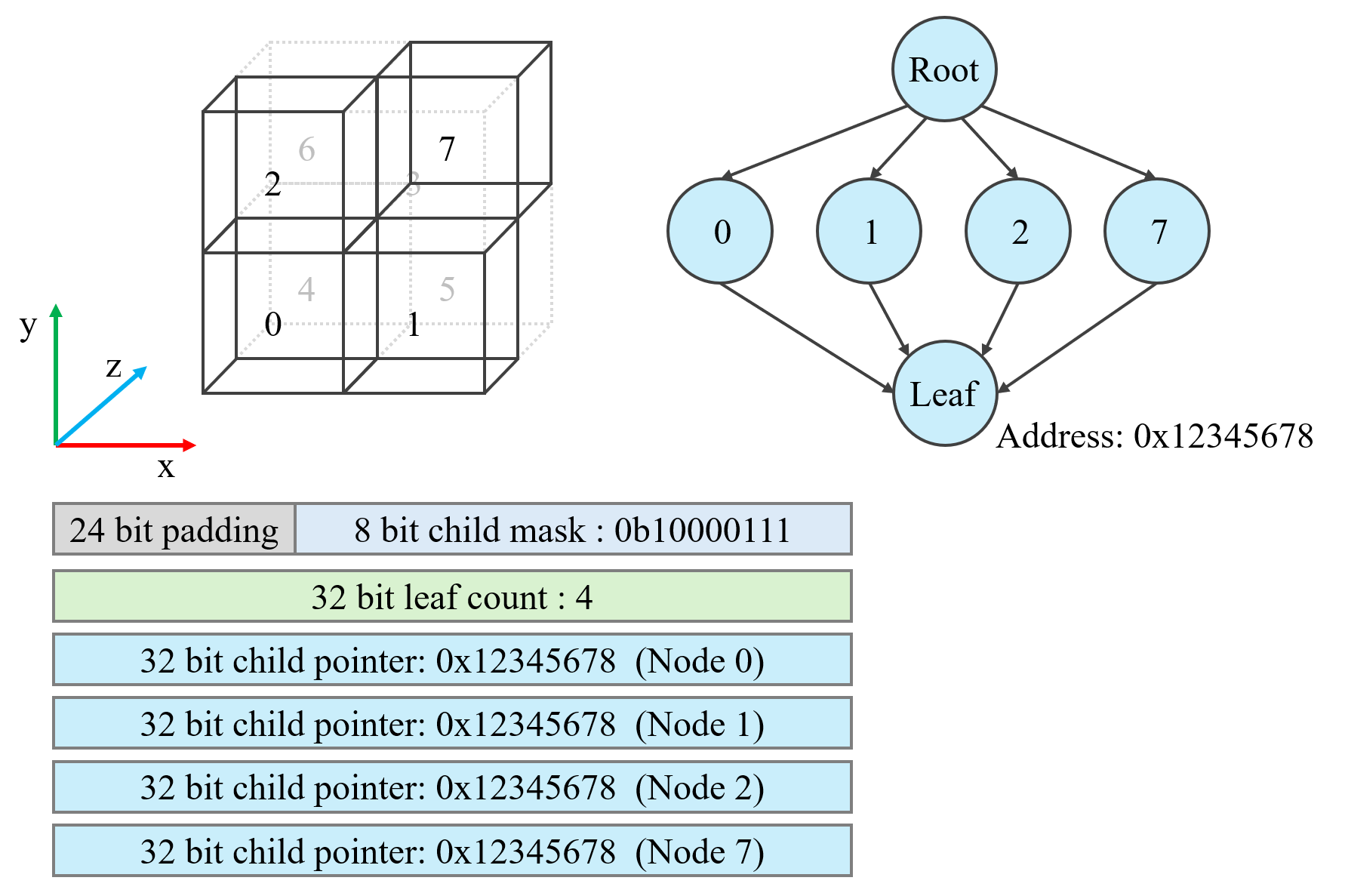}
    \caption{Node structure example. \textbf{Top left:} A voxel region of size $2\times 2\times 2$ containing 4 voxels. \textbf{Top right:} The SVDAG formed by this voxel region. \textbf{Bottom:} The structure of the root node representing the SVDAG for this region. 
    We use a 64-bit bitmap to represent the geometric information of the deepest $4\times 4\times 4$ sub-chunks.}
    \Description{Node Structure}
    \label{fig:node_structure}
\end{figure}}{\begin{figure}[ht]
  \centering
  \begin{minipage}[b]{0.45\textwidth}
    \includegraphics[width=\textwidth]{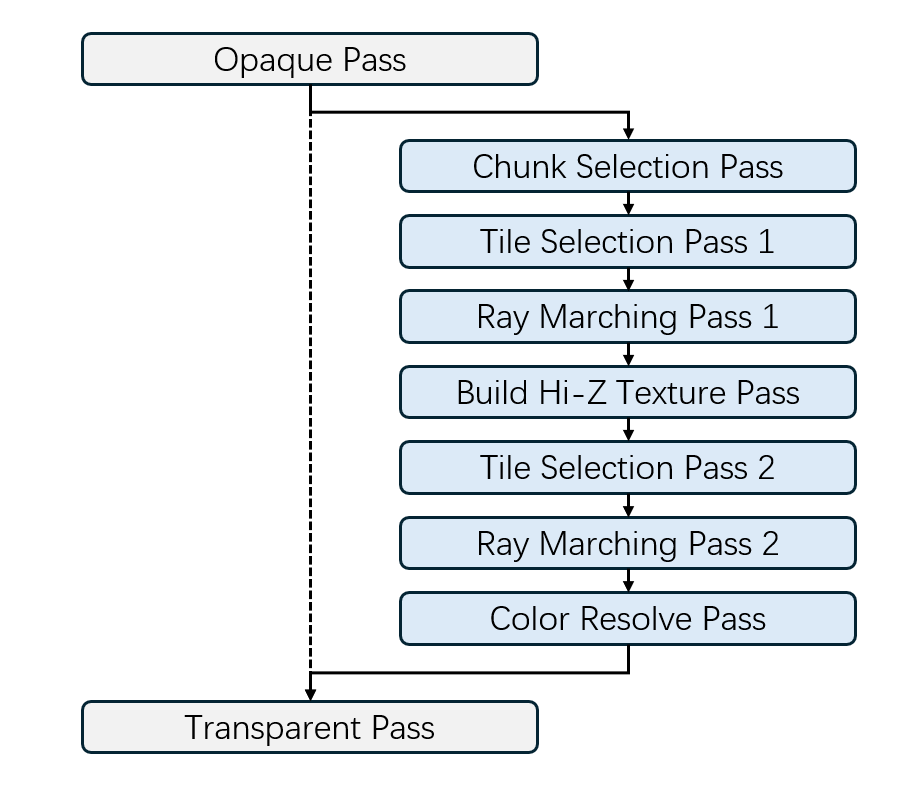}
    \caption{The rendering pipeline overview. Our voxel rendering passes are inserted between the opaque pass and transparent pass of the forward rendering pipeline.}
    \label{fig:pipeline}
  \end{minipage}
  \hfill
  \begin{minipage}[b]{0.5\textwidth}
    \includegraphics[width=\textwidth]{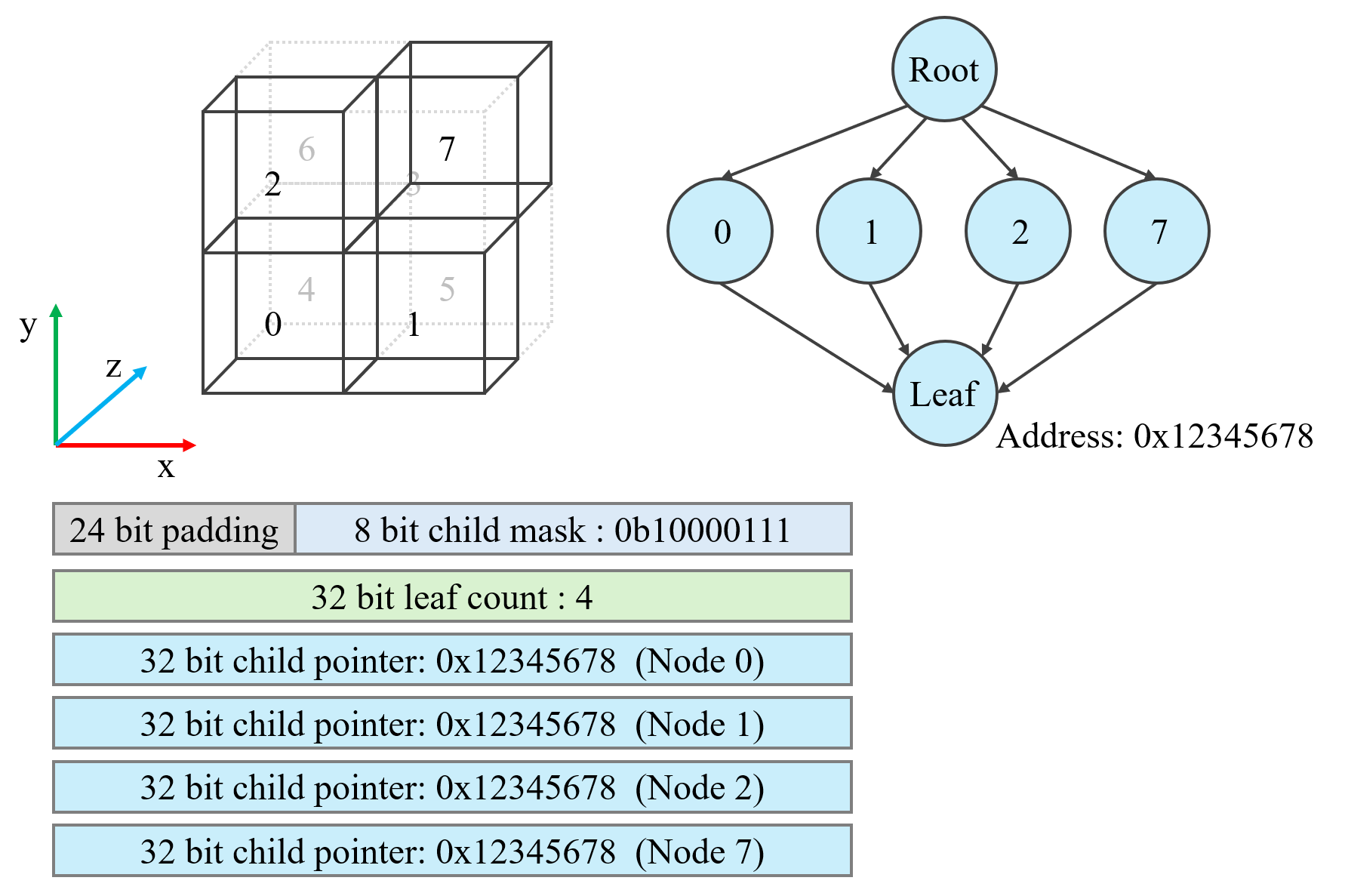}
    \caption{Node structure example. \textbf{Top left:} A voxel region of size $2\times 2\times 2$ containing 4 voxels. \textbf{Top right:} The SVDAG formed by this voxel region. \textbf{Bottom:} The structure of the root node representing the SVDAG for this region. 
    We use a 64-bit bitmap to represent the geometric information of the deepest $4\times 4\times 4$ sub-chunks.}
    \Description{Node Structure}
    \label{fig:node_structure}
  \end{minipage}
\end{figure}}

\subsection{Sparse Voxel DAG Chunk}

Due to the pointer-linked structure of SVDAG, it is not cache-friendly. 
As the voxel resolution continues to increase, the depth of the SVDAG also increases, leading to more indirect jumps along the path from the root node to the leaf nodes, which results in memory access bottlenecks. 
To reduce the number of indirect jumps, we propose that instead of using a single SVDAG with a large depth, multiple SVDAGs with smaller depths should be used to represent the entire scene. 
We divide the entire open-world game map into a series of axis-aligned cubic regions, each sized $M^3$ , with each region maintaining a chunk of resolution $256^3$ voxels. For each chunk, we employ Sparse Voxel DAG for compression and representation. 
Ultimately, the entire open-world voxel map is segmented into several Sparse Voxel DAGs. 
Each chunk contains two types of information: geometric information indicating whether a voxel exists at the position $(x, y, z)$, and color information specifying the color of the voxel at that position. 
We decouple the geometric and color information of the chunk and compress them separately.

\subsection{Geometry Compression}

For the compression of geometric data, we employ a method similar to SVDAG, which involves merging isomorphic subtrees of the Sparse Voxel Octree to form a DAG. 
The spatial location information of nodes in the original Octree is implicitly encoded in the path from the source of the DAG to its subsequent nodes, allowing ray marching on the compressed DAG to be performed similarly to that on the Octree. 
A node in the DAG is represented using 8 to 40 bytes. The first 4 bytes of each node's data contain a 32-bit head node, where the lowest 8 bits represent a child mask that indicates the presence of corresponding child nodes. 
The following 4 bytes store the number of leaf nodes in the subtree rooted at that node. 
The final 0 to 32 bytes contain pointers to child nodes, with the number of pointers corresponding to the number of binary 1s in the child mask. 
For the three deepest layers of nodes, we consider them as individual $4\times 4\times 4$ \textit{leaf chunks} and utilize a 64-bit bitmap to represent the geometric information of the leaf chunks.
Figure~\ref{fig:node_structure} illustrates an example of a node.

\subsection{Color Compression}

We further decouple geometric and color data~\cite{dolonius2017compressing} for the compression of the color information.
Each node in the SVDAG records the number of leaf nodes in the subtree rooted at that node. 
By accumulating the counts of child nodes along the path from the root node to the leaf node, we can determine the index of the current leaf node in the depth-first search (DFS) order. 
This index is then used to store the color of the leaf node in a separate, independent color array, which is subsequently compressed.

Since the compression of color data and the extraction of colors from the compressed data are independent steps from ray marching and are not the primary focus of our work, and given that our LOD mechanism and streaming loading mechanism significantly reduce memory usage, a higher compression ratio does not yield substantial improvements. 
For simplicity, we made slight modifications to the Dolonius et al.~\cite{dolonius2017compressing} method. 
We compress the color array into several blocks, where each block represents a contiguous segment of the original color array. 
Each block records the starting position of this segment in the original color array, denoted as $block.start\_index$, the length of the contiguous segment as $block.length$, and the average color of this segment as $block.color$. We set a tolerable error threshold $e_c$, and in our implementation, we set $e_c = 0.05$. 
For a color $c$ of a leaf node adjacent to the current block in the original color array, if $|| c - block.color || \leq e_c$, the leaf node can be merged with the current block, updating $block.color$ to the current average color, which is then treated as the color of that leaf node. 
Assuming the length of the color array is $n$, the time complexity of this algorithm is $O(n)$.

\subsection{Level of Details and Streaming}

To reduce VRAM usage, we implemented a LOD mechanism for the chunks. 
We consider the smallest voxel chunk, which maintains a resolution of $M$ for a region of size $M^3$, as LOD 0.
Using a method similar to an octree, we aggregate eight LOD 0 chunks that maintain a region of size $(2M)^3$ into a single LOD 1 chunk, which retains the same resolution of $256^3$. 
This process continues recursively. Thus, each voxel in an LOD $n$ chunk is derived from the aggregation of eight voxels from the LOD $n-1$ chunks. 
If the number of non-empty voxels among these eight is greater than or equal to a predefined $density$, we create a new voxel and use the average color of these voxels as the color of the aggregated voxel. 
In our implementation, we set $density=2$.

\begin{figure}[h]
    \centering
    \includegraphics[width=1.0\textwidth]{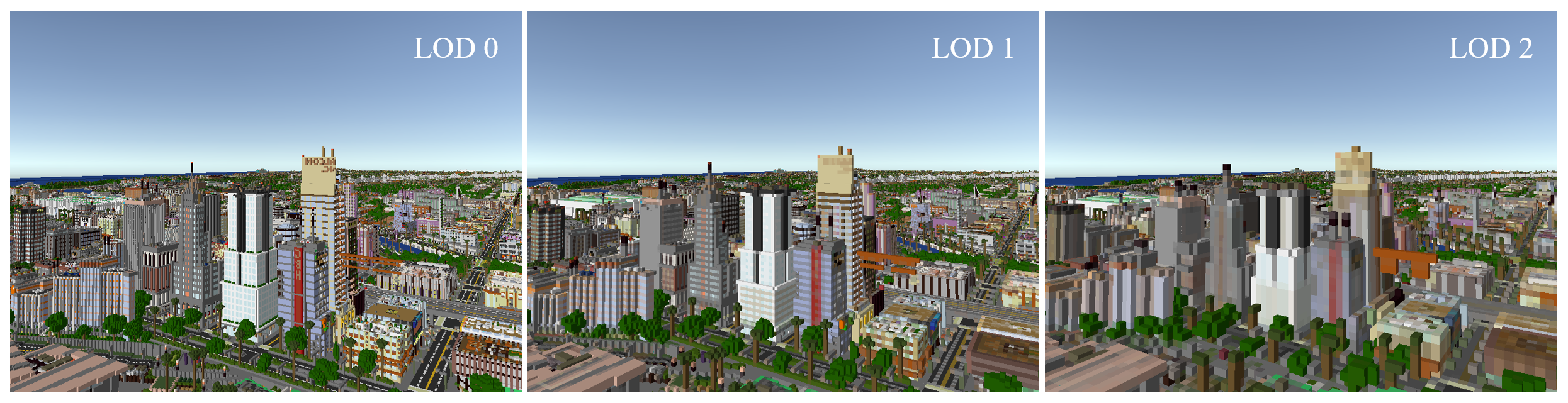}
    \caption{
    A visualization of different LODs. 
    Taking the building located at the center of the image as an example, more details can be observed at lower-level LODs.
    }
    \Description{
    LOD Example. Please observe the building located at the center of the image.
    }
\end{figure}

We maintain an implicit octree on the CPU side to organize the chunks and filter those that need to be loaded. 
Whenever the chunk containing the camera changes, we initiate a coroutine to perform a recursive search starting from the root node of the octree. 
Each time a node in the octree is accessed, we first check whether the region it manages contains LOD chunks. 
If it does not, we continue the recursion downward. If the current node does contain LOD chunks, we use the formula (Equation 1) from Virtual Voxel to evaluate the LOD error based on the position and size of the chunk as well as the position of the camera. 
The StreamingFactor is a predefined parameter; the larger its value, the finer the details at greater distances, which correspondingly requires more VRAM. 
If the LOD error is greater than 0, further subdivision is necessary. 
If the LOD error is less than 0, the chunk is marked as needing to be loaded. When determining the loading order of chunks, we can prioritize loading the chunks with higher LOD levels, allowing us to directly remove all child blocks associated with that chunk to free up VRAM. When there is sufficient space in the VRAM, chunks that are closer to the player are prioritized for loading.

\begin{equation}
LODError = (ChunkSize \times StreamingFactor) - ||ChunkCenterPos - CameraPos||
\end{equation}

\begin{figure}[h]
    \centering
    \includegraphics[width=0.8\textwidth]{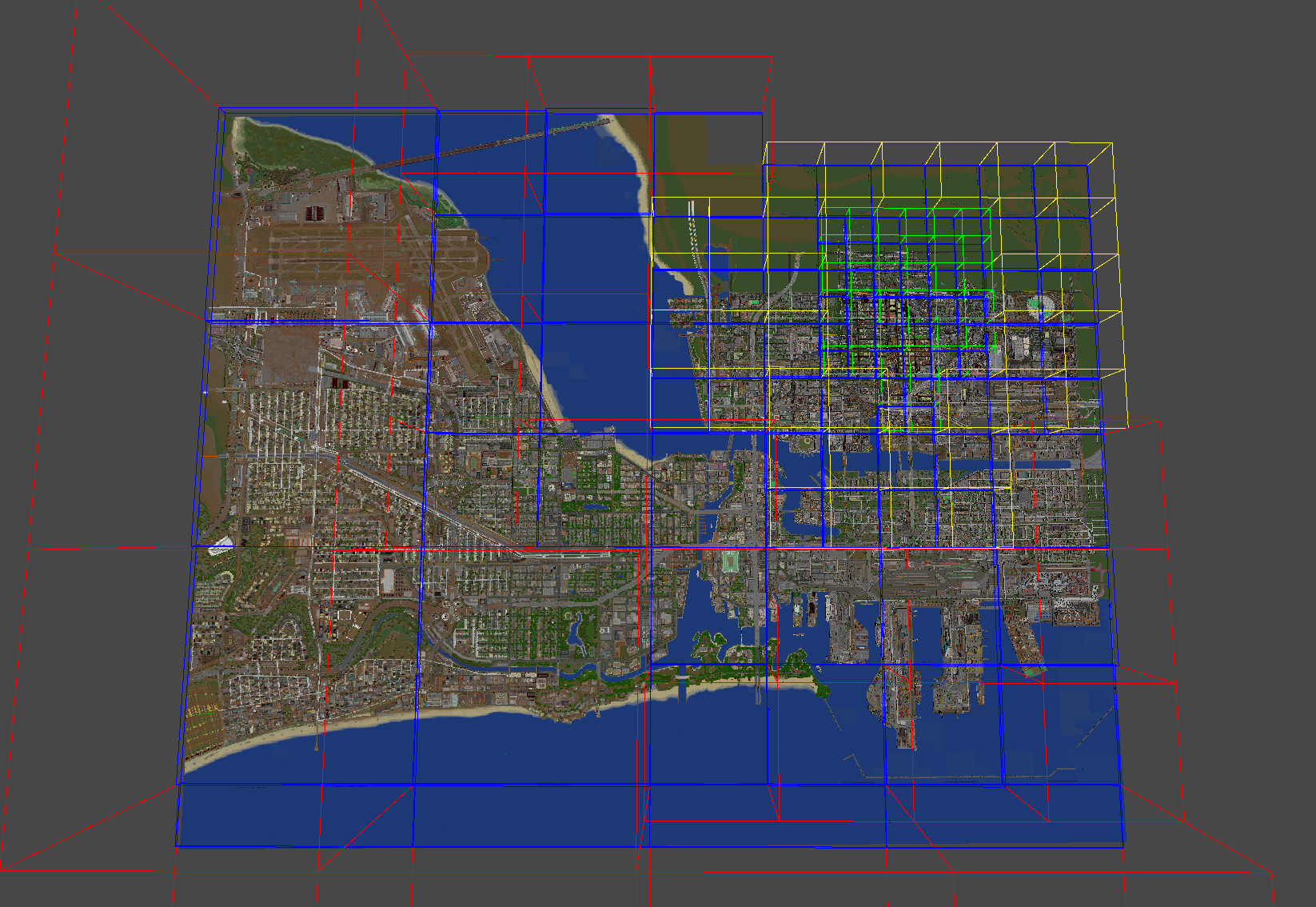}
    \caption{The camera is located at the green outline in the upper right corner of the image, and the chunks in the world are loaded on demand by selecting the appropriate LOD level.}
    \Description{World Partition}
\end{figure}

\subsection{GPU-Driven Voxel Rendering Pipeline}

We referenced the approach shared by Ubisoft in the GPU-Driven rendering pipelines~\cite{haar2015gpu} and proposed the GPU-Driven Voxel Rendering Pipeline, where most operations are performed on the GPU side. 
To reduce the number of indirect jumps, we use multiple shallow SVDAGs to organize the scene. However, if the rendering scheme for the SVDAGs is not carefully designed, it may not only fail to reduce the number of indirect jumps but also lead to significant overdraw. 
We designed a GPU-Driven rendering pipeline for efficient voxel rendering. 
The entire rendering pipeline mainly includes several passes: chunk selection pass, tile selection pass, ray marching pass, and build Hi-Z pass. 
All of these passes are executed in compute shaders. 
Our custom passes are inserted after the opaque pass and before the transparent pass in the forward rendering pipeline. 
Before all our passes begin, all opaque objects in the scene have already been rendered. 
We then blit the current color target and depth target of the camera into two new textures before starting our passes.

We precomputed the axis-aligned bounding boxes for the portions of each chunk that contain valid voxels, and in the chunk selection pass, we used frustum culling based on the bounding boxes of the chunks to eliminate those that are currently not visible to the camera.

Due to the high cost of ray intersection with chunks represented by SVDAG, we aim to minimize ray and chunk intersections to reduce overdraw and ensure rendering efficiency. 
In the tile selection pass, we divide the screen into a series of tiles with $8\times 8$ pixels. For every $4\times 4$ tiles, we allocate a thread group, assigning one thread to each tile. 
We project screen rays from each tile to determine which chunks that passed the chunk selection pass contribute to the current tile. 
We refer to the chunk currently containing the camera as the \textit{self chunk}. 
For chunks other than the self chunk, we do not perform intersections with the SVDAG but instead with the axis-aligned bounding boxes. 
Since the camera is guaranteed to be outside the axis-aligned bounding box of chunks other than the self chunk, the depth value of the intersection point obtained with the bounding box will always be less than the depth value obtained from the SVDAG. 
We use this intersection depth value as the depth value for the current tile. 
We referenced the Hierarchical Z-Buffer proposed by N. Greene et al.~\cite{greene1993hierarchical}, using Hi-Z Textures for Hi-Z occlusion culling of the current tile. 
This approach poses no issues for Hi-Z occlusion culling, as using a depth value smaller than the actual depth value results in more conservative culling, ensuring that no tiles are incorrectly culled. 
However, for the self chunk, since the camera is within the self chunk, the depth value of the intersection between the ray and the axis-aligned bounding box will always be greater than the actual depth value obtained from the SVDAG, leading to incorrect culling in Hi-Z Occlusion Culling. Therefore, for the self chunk, we need to perform intersections with the SVDAG. 
After the tile selection pass, we identify which chunks contribute to the corresponding tiles, ultimately resulting in a series of Tile-Chunk pairs. 
We use a custom TileInfo structure buffer to store the coordinates of each tile that passed the intersection test and the corresponding chunk ID.

\begin{figure}[h]
    \centering
    \includegraphics[width=0.9\textwidth]{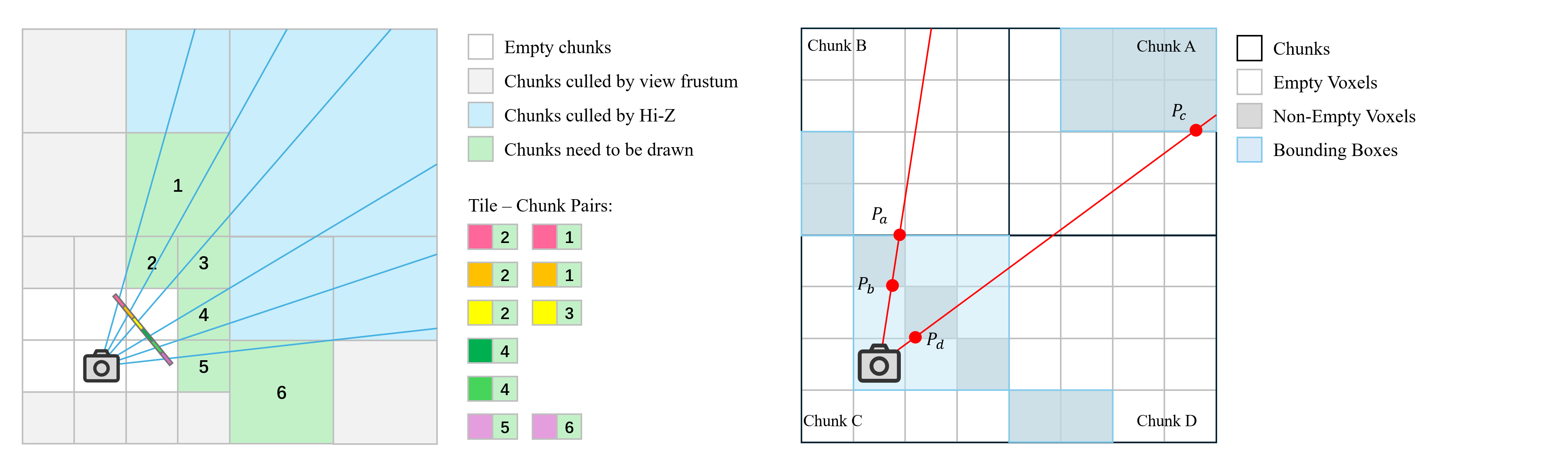}
    \caption{\textbf{Left:} A large number of chunks will be culled by frustum culling and Hi-Z occlusion culling, resulting in a series of Tile-Chunk pairs. \textbf{Right: } The camera is located in chunk $C$, projecting two rays. The first ray intersects with chunk $C$ itself, while the second ray intersects with both chunk $C$ and chunk $A$. For the Tile-Chunk pairs associated with chunk C, it is necessary to use the depth values at the intersection points with the SVDAG, such as $P_a, P_b$, as the depth values for the tiles to prevent incorrect Hi-Z culling of the tiles.}
    \Description{The example of occlusion culling.}
\end{figure}

For all Tile-Chunk pairs filtered in the tile selection pass, we perform an indirect dispatch compute shader to initiate the DAG ray marching pass, assigning a thread group to each tile. 
We referenced the method by Burns et al.~\cite{burns2013visibility} and used a 64-bit visibility buffer to record visibility information. 
In this pass, we employ a ray intersection method similar to that in ESVO to calculate the intersection points between each pixel on every tile and the corresponding chunk's SVDAG. We use InterlockedMax() to write the depth values, normal information, Chunk IDs, and voxel coordinates at the intersection points into a 64-bit visibility buffer that matches the size of the screen. 
Following the approach by M. Sch{\"u}tz et al.~\cite{schutz2022software}, we store the depth values in the upper 24 bits of the visibility buffer, setting the near plane to use $2^{24} - 1$ as the depth value and the far plane to use $0$ as the depth value. 
Using InterlockedMax() to write intersection information into the visibility buffer effectively simulates depth testing. 
Since we assume that voxels are axis-aligned, we only need 3 bits to store the normals of the intersecting voxel surfaces. The detailed bit layout of the Visibility Buffer is shown in Figure 7.

In the Color Resolve Pass, we divide the screen into $8\times 8$ tiles and allocate a thread group to each tile in the compute shader, assigning one thread to each pixel to compute its color. 
First, we retrieve the information for the pixel from the visibility buffer. 
If the information is valid, we extract the normal, chunk ID, and voxel position. 
We then calculate the DFS order of the intersection node in the SVDAG based on the voxel position. 
Using the DFS order, we perform a binary search in the color block array to find the corresponding color block. Finally, we shade based on the normal and color, writing the shading result into the color texture.

\begin{figure}[h]
    \centering
    \includegraphics[width=0.8\textwidth]{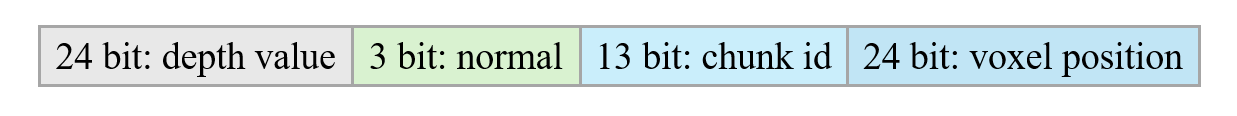}
    \caption{The bit layout of the 64-bit visibility buffer is as follows: the highest 24 bits store the depth value of the intersection point, the next 3 bits store the axis-aligned normal, the following 13 bits store the chunk ID, and the final 24 bits store the voxel's X, Y, and Z coordinates.}
    \Description{The bit layout of visibility buffer.}
\end{figure}

In the build Hi-Z pass, we construct the Hi-Z Texture from the depth information of the current frame's Depth Texture. 
Since the camera may be moving at high speeds, using the previous frame's Hi-Z Texture for tile culling may not be accurate and could lead to incorrect culling. 
Therefore, we re-execute the tile selection pass, performing an indirect dispatch compute shader on all tiles that were culled in the first tile selection pass. 
We use the current frame's Hi-Z Texture to reapply Hi-Z occlusion culling, filtering out all tiles that were incorrectly culled. 
We then perform DAG Ray Marching on these tiles again, writing the intersection colors and depth information into the texture.

Finally, we blit the Depth Texture and Color Texture back to the depth target and color target, completing our voxel rendering process. We then proceed to the Transparent Pass to render the transparent meshes.

\section{Experiments and Discussions}

\begin{figure}[h]
    \centering
    \includegraphics[width=1.0\textwidth]{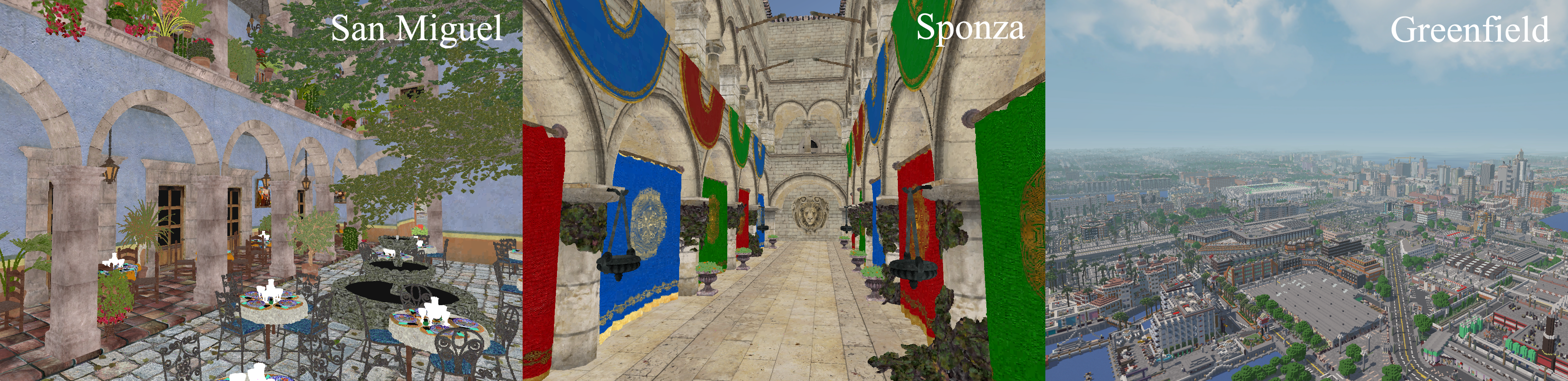}
    \caption{
    A visualization of the three test scenes, i.e., San Miguel, Sponza, and Greenfield, in our experiments.
    }
    \Description{We use the San Miguel, Sponza, and Greenfield as test scenes.}
\end{figure}

We implemented our method using the RenderGraph API in the Unity 6 game engine, with Vulkan as the rendering backend, the shader compiler set to dxc, and using il2cpp for building. We ran measurements on an AMD Ryzen 5 5600X CPU with 64GB of DDR4 memory and an NVIDIA GeForce RTX 3060Ti GPU under Windows.

We conducted experiments on the San Miguel, Sponza, and Greenfield scenes. San Miguel and Sponza are classic graphics test scenes. San Miguel features a large amount of complex and discontinuous geometry (such as leaves), which can roughly reflect the performance of our method on natural scenes, while the geometry in Sponza is somewhat simpler. We voxelized San Miguel and Sponza into a series of datasets with resolutions ranging from 8K to 64K. Greenfield is derived from a famous Minecraft save file and represents a real game scene with a large-scale modern city, allowing us to evaluate our method's performance in real open-world games. One cube in Minecraft corresponds to one voxel in our implementation, so for the Greenfield scene, we are unable to construct datasets at resolutions other than 8K. We randomly selected 10 camera positions in the scenes as sample points and set the $StreamingFactor$ to $2.0$. 

\subsection{Rendering Time}
\begin{figure}[h]
    \centering
    \includegraphics[width=1.0\textwidth]{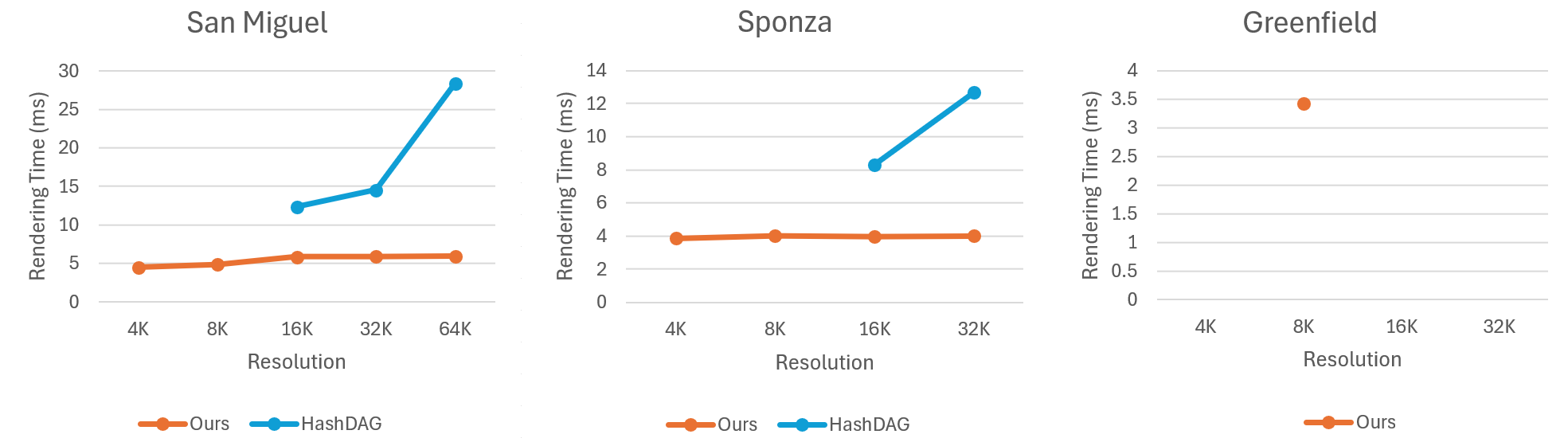}
    \caption{Compare the rendering time of our method and other methods for a single frame. The rendering speed of our method is faster than that of HashDAG, and our method's rendering speed is less sensitive to scene resolution.}
    \Description{Comparison of Rendering Times}
    \label{fig:rendering_time_cmp}
\end{figure}

We first measured the time required to complete the rendering of a single frame by comparing our method with SVDAG and HashDAG in Figure \ref{fig:rendering_time_cmp}. 
We found that our method outperformed HashDAG in rendering time, and as the voxel scale increased, the increase in rendering time was relatively smooth. 
At a voxel resolution of $64k$, the number of voxels reached the ten billion scale, and at this scale, our method completed rendering in an average of 6 milliseconds, demonstrating that our approach can achieve real-time rendering of large-scale voxels. 
For voxel scenes with a resolution above 32K, our method achieves an average rendering speed that is 2 to 4 times faster than HashDAG.

\begin{figure}[h]
    \centering
    \includegraphics[width=1.0\textwidth]{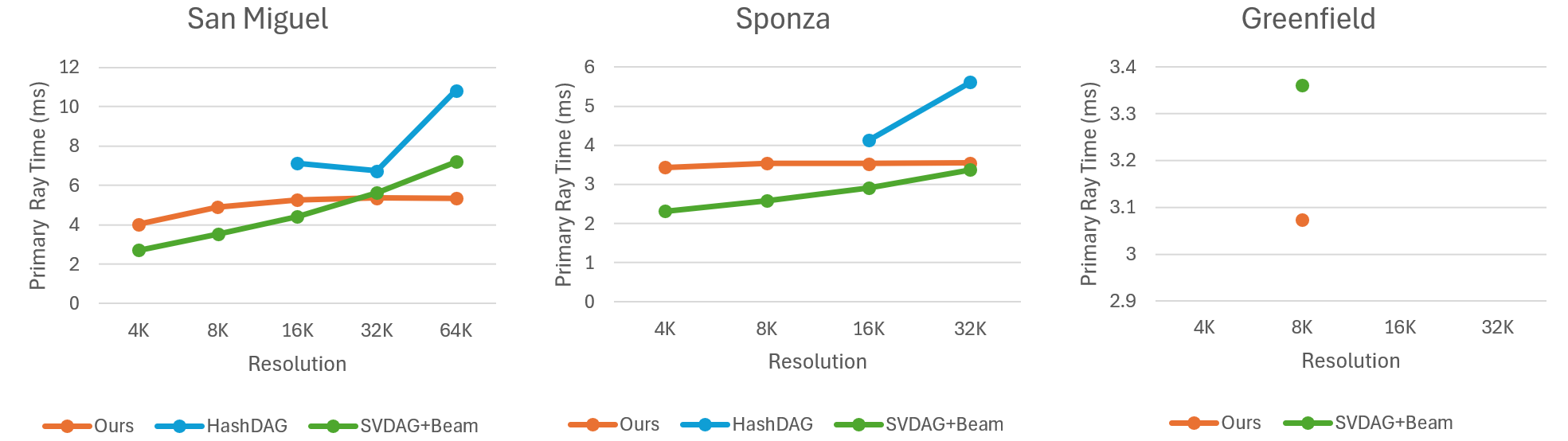}
    \caption{Comparison of the time to compute primary ray intersection points. Our method has advantages in high-resolution scenes and real open-world scenes.}
    \Description{Comparison of the Time to Compute Primary Ray Intersection Points}
    \label{fig:primary_ray_time_cmp}
\end{figure}

In Figure \ref{fig:primary_ray_time_cmp}, we tested the time taken to compute the intersection points of the primary rays. 
In addition to comparing with SVDAG, we included a comparison with a single SVDAG utilizing beam optimization~\cite{laine2010efficient}
as part of our ablation study.
Since color computation is not considered at this stage, our method is relatively close to the beam-optimized SVDAG, but it slightly outperforms it in scenarios with higher resolutions. 
This is because scenes with larger resolutions lead to deeper SVDAG structures, which are less cache-friendly, while our method uses multiple SVDAGs with smaller depths, reducing the number of indirect jumps during queries.

\subsection{Memory Efficiency}
We also compared our method with HashDAG and SVDAG in terms of memory efficiency. Since our method utilizes a memory pool and incorporates an LOD streaming mechanism, we recorded the maximum occupancy of the memory pool during random walks in the test scenes as the memory requirement in Table \ref{tab:memory_cmp}. 
HashDAG also designed a memory pool, but they need to load the complete scene data into memory, so we recorded the size of the scene files instead of the memory pool size. 
For the implementation of the SVDAG method, we directly merged all LOD 0 SVDAG chunks of our method into a single SVDAG, resulting in a slightly larger storage size than the original implementation. 
For each LOD level, we recorded the number of chunks in Table \ref{tab:chunk_count}, the number of nodes in Table \ref{tab:node_count}, and the storage size of the geometry in Table \ref{tab:lod_geometry_storage_size}.
Our method requires more disk space than HashDAG and SVDAG because it needs to store multiple levels of LOD structures, and we did not merge subtrees with depths greater than 8 when constructing SVDAG chunks. Additionally, we stored the count of leaf nodes in the subtree used for color lookup.
Our method also requires more disk space for color information because color compression is not the focus of our work, and we opted for a simpler color compression method. While we could certainly choose more complex color compression techniques, in actual gameplay, the voxel values may not only represent color but can also have significance in gameplay (consider Minecraft). This means that compression for color (or attributes) cannot utilize complex lossy compression algorithms, especially for LOD 0 chunks.
However, due to our LOD streaming mechanism, our method leads in memory usage, only about $5\%$ of the complete voxel scene data needs to be loaded into VRAM when navigating the scene, allowing it to meet memory requirements even on low-end hardware.

\begin{table*}
\caption{Comparison of the storage sizes in scenes with different resolutions. Our method has a lower VRAM usage compared to other methods.}
\label{tab:memory_cmp}
\vspace*{-6pt}
\resizebox{\textwidth}{!}{
\begin{tabular}{cccccccccc}
\toprule
\textbf{Scene} & \textbf{Method} & \textbf{Resolution} & \textbf{Geometry (MB)} & \textbf{Color (MB)} & \textbf{Disk (MB)} & \textbf{VRAM (MB)} & \textbf{VRAM/Disk} & \textbf{Max Active Chunks} & \textbf{Ours/HashDAG (VRAM)} \\ \midrule
San Miguel     & \textbf{Ours}   & 64K                 & 2217.07                & 21016.02            & 23233.09           & 424.11             & 2\%                & 975                        & 12\%                         \\
San Miguel     & \textbf{Ours}   & 32K                 & 645.69                 & 7220.57             & 7866.26            & 385.95             & 5\%                & 730                        & 34\%                         \\
San Miguel     & \textbf{Ours}   & 16K                 & 185.21                 & 2311.40             & 2496.61            & 292.86             & 12\%               & 575                        & 83\%                         \\ \cline{1-10}
San Miguel     & HashDAG         & 64K                 & 639.86                 & 2835.19             & 3475.05            & 3475.05            & 100\%              & /                          & /                            \\
San Miguel     & HashDAG         & 32K                 & 221.50                 & 906.35              & 1127.85            & 1127.85            & 100\%              & /                          & /                            \\
San Miguel     & HashDAG         & 16K                 & 72.04                  & 280.30              & 352.34             & 352.34             & 100\%              & /                          & /                            \\ \cline{1-10}
San Miguel     & SVDAG           & 64K                 & 2483.90                & /                   & 2483.90            & 2483.90            & 100\%              & /                          & /                            \\
San Miguel     & SVDAG           & 32K                 & 795.42                 & /                   & 795.42             & 795.42             & 100\%              & /                          & /                            \\
San Miguel     & SVDAG           & 16K                 & 242.57                 & /                   & 242.57             & 242.57             & 100\%              & /                          & /                            \\ \cline{1-10}
Sponza         & \textbf{Ours}   & 32K                 & 806.70                 & 19708.57            & 20515.27           & 327.68             & 2\%                & 636                        & 11\%                         \\
Sponza         & \textbf{Ours}   & 16K                 & 233.62                 & 8122.35             & 8355.97            & 307.18             & 4\%                & 556                        & 32\%                         \\ \cline{1-10}
Sponza         & HashDAG         & 32K                 & 172.33                 & 2830.55             & 3002.88            & 3002.88            & 100\%              & /                          & /                            \\
Sponza         & HashDAG         & 16K                 & 64.93                  & 902.02              & 966.95             & 966.95             & 100\%              & /                          & /                            \\ \cline{1-10}
Sponza         & SVDAG           & 32K                 & 743.47                 & /                   & 743.47             & 743.47             & 100\%              & /                          & /                            \\
Sponza         & SVDAG           & 16K                 & 230.04                 & /                   & 230.04             & 230.04             & 100\%              & /                          & /                            \\ \cline{1-10}
Greenfield     & \textbf{Ours}   & 8K                  & 58.85                  & 779.54              & 838.38             & 140.8              & 17\%               & /                          & /                            \\ \cline{1-10}
Greenfield     & SVDAG           & 8K                  & 46.14                  & /                   & 46.14              & 46.14              & 100\%              & /                          & /                            \\ \bottomrule
\end{tabular}
}
\end{table*}

\begin{table*}
\caption{he number of chunks per LOD in our method.}
\label{tab:chunk_count}
\vspace*{-6pt}
\resizebox{\textwidth}{!}{
\begin{tabular}{cccccccccccc}
\toprule
\textbf{Scene} & \textbf{Resolution} & \textbf{LOD 0} & \textbf{LOD 1} & \textbf{LOD 2} & \textbf{LOD 3} & \textbf{LOD 4} & \textbf{LOD 5} & \textbf{LOD 6}                               & \textbf{LOD 7}                               & \textbf{LOD 8}                               & \textbf{Total} \\ \midrule
San Miguel     & 4K                  & 248            & 48             & 8              & 4              & 1              &                &                                              &                                              &                                              & 309            \\
San Miguel     & 8K                  & 2310           & 248            & 48             & 8              & 2              & 1              &                                              &                                              &                                              & 2617           \\
San Miguel     & 16K                 & 6306           & 1340           & 248            & 48             & 8              & 2              & 1                                            &                                              &                                              & 7953           \\
San Miguel     & 32K                 & 32131          & 6304           & 1340           & 248            & 48             & 8              & 2                                            & 1                                            &                                              & 40082          \\
San Miguel     & 64K                 & 163618         & 32041          & 6293           & 1340           & 248            & 48             & 8                                            & 2                                            & 1                                            & 203599         \\ \cline{1-12}
Sponza         & 4K                  & 1086           & 160            & 24             & 4              & 1              &                &                                              &                                              &                                              & 1275           \\
Sponza         & 8K                  & 5406           & 1085           & 160            & 24             & 4              & 1              &                                              &                                              &                                              & 6680           \\
Sponza         & 16K                 & 25129          & 5405           & 1085           & 160            & 24             & 4              & 1                                            &                                              &                                              & 31808          \\
Sponza         & 32K                 & 112528         & 25128          & 5404           & 1085           & 160            & 24             & 4                                            & 1                                            &                                              & 144334         \\ \cline{1-12}
Greenfield     & 8K                  & 856            & 201            & 56             & 16             & 4              & 1              &  &  &  & 1134           \\ \bottomrule
\end{tabular}
}
\end{table*}

\begin{table*}
\caption{The number of nodes per LOD in our method.}
\label{tab:node_count}
\vspace*{-6pt}
\resizebox{\textwidth}{!}{
\begin{tabular}{cccccccccccc}
\toprule
\textbf{Scene} & \textbf{Resolution} & \textbf{LOD 0} & \textbf{LOD 1} & \textbf{LOD 2} & \textbf{LOD 3} & \textbf{LOD 4} & \textbf{LOD 5} & \textbf{LOD 6} & \textbf{LOD 7} & \textbf{LOD 8} & \textbf{Total} \\ \midrule
San Miguel     & 4K                  & 867693         & 197737         & 45419          & 11174          & 2651           &                &                &                &                & 1124674        \\
San Miguel     & 8K                  & 2996001        & 764995         & 182477         & 43445          & 10895          & 2601           &                &                &                & 4000414        \\
San Miguel     & 16K                 & 10076072       & 2795518        & 725116         & 176960         & 42684          & 10743          & 2579           &                &                & 13829672       \\
San Miguel     & 32K                 & 33222059       & 10008118       & 2706180        & 709476         & 174649         & 42198          & 10696          & 2571           &                & 46875947       \\
San Miguel     & 64K                 & 104092037      & 34590631       & 9811871        & 2667878        & 701588         & 173396         & 42002          & 10661          & 2565           & 152092629      \\ \cline{1-12}
Sponza         & 4K                  & 949668         & 283452         & 76500          & 19843          & 5152           &                &                &                &                & 1334615        \\
Sponza         & 8K                  & 2964837        & 960481         & 271686         & 75340          & 19273          & 5040           &                &                &                & 4296657        \\
Sponza         & 16K                 & 9214478        & 3159527        & 930941         & 268763         & 74151          & 19053          & 5048           &                &                & 13671961       \\
Sponza         & 32K                 & 30266834       & 10232722       & 3092768        & 925400         & 263966         & 73609          & 18991          & 5035           &                & 44879325       \\ \cline{1-12}
Greenfield     & 8K                  & 3167715        & 683511         & 141683         & 29997          & 7069           & 601            &                &                &                & 4030576        \\ \bottomrule
\end{tabular}
}
\end{table*}

\begin{table*}
\caption{The geometry storage size per LOD in our method.}
\label{tab:lod_geometry_storage_size}
\vspace*{-6pt}
\resizebox{\textwidth}{!}{
\begin{tabular}{cccccccccccc}
\toprule
\textbf{Scene} & \textbf{Resolution} & \textbf{LOD 0} & \textbf{LOD 1} & \textbf{LOD 2} & \textbf{LOD 3} & \textbf{LOD 4} & \textbf{LOD 5} & \textbf{LOD 6}                               & \textbf{LOD 7}                               & \textbf{LOD 8}                               & \textbf{Total} \\ \midrule
San Miguel     & 4K                  & 11.27          & 2.61           & 0.62           & 0.15           & 0.04           &                &                                              &                                              &                                              & 14.69          \\
San Miguel     & 8K                  & 39.52          & 10.05          & 2.43           & 0.59           & 0.15           & 0.04           &                                              &                                              &                                              & 52.79          \\
San Miguel     & 16K                 & 135.62         & 36.87          & 9.59           & 2.37           & 0.58           & 0.15           & 0.04                                         &                                              &                                              & 185.21         \\
San Miguel     & 32K                 & 464.24         & 133.18         & 35.76          & 9.40           & 2.34           & 0.58           & 0.15                                         & 0.04                                         &                                              & 645.69         \\
San Miguel     & 64K                 & 1564.00        & 473.69         & 130.71         & 35.29          & 9.30           & 2.32           & 0.57                                         & 0.15                                         & 0.04                                         & 2217.07        \\ \cline{1-12}
Sponza         & 4K                  & 14.99          & 4.29           & 1.15           & 0.30           & 0.08           &                &                                              &                                              &                                              & 20.81          \\
Sponza         & 8K                  & 48.96          & 15.09          & 4.17           & 1.14           & 0.29           & 0.08           &                                              &                                              &                                              & 69.74          \\
Sponza         & 16K                 & 161.75         & 51.43          & 14.80          & 4.14           & 1.13           & 0.29           & 0.08                                         &                                              &                                              & 233.62         \\
Sponza         & 32K                 & 559.73         & 175.91         & 50.75          & 14.72          & 4.09           & 1.12           & 0.29                                         & 0.08                                         &                                              & 806.70         \\ \cline{1-12}
Greenfield     & 8K                  & 46.17          & 10.17          & 2.02           & 0.39           & 0.09           & 0.01           &  &  &  & 58.85          \\ \bottomrule
\end{tabular}
}
\end{table*}

\subsection{Rendering Quality}

We also conducted experiments on rendering quality under different $StreamingFactor$ values. 
We used the rendering results of the scene when all LOD 0 chunks were loaded as the ground truth.
We randomly selected 10 camera positions as sample points and recorded the average SSIM for $StreamingFactor$ values ranging from $1.2$ to $2.4$, along with the maximum number of loaded chunks while traveling through the scene, as shown in Table \ref{tab:ssim_table}.
The experimental results indicate that when the $StreamingFactor$ is $2.0$, the SSIM is near 0.9, and it maintains relatively high rendering quality while loading fewer chunks. 
If our method is used on low-end hardware or if there is a desire to reduce the memory usage of our method, $StreamingFactor$ can be appropriately lowered.

\begin{table*}
\caption{Our method's SSIM and Max Active Chunk Count at different $StreamingFactors$.}
\label{tab:ssim_table}
\begin{tabular}{ccccc}
\toprule
\textbf{Scene} & \textbf{Resolution} & \textbf{StreamingFactor} & \textbf{Average SSIM} & \textbf{Max Active Chunks} \\ \midrule
San Miguel     & 8K                  & 2.4                      & 0.924                 & 470                        \\
San Miguel     & 8K                  & 2.0                      & 0.888                 & 341                        \\
San Miguel     & 8K                  & 1.6                      & 0.845                 & 236                        \\
San Miguel     & 8K                  & 1.2                      & 0.694                 & 118                        \\ \cline{1-5}
Sponza         & 8K                  & 2.4                      & 0.86                  & 731                        \\
Sponza         & 8K                  & 2.0                      & 0.8                   & 473                        \\
Sponza         & 8K                  & 1.6                      & 0.703                 & 300                        \\
Sponza         & 8K                  & 1.2                      & 0.587                 & 165                        \\ \cline{1-5}
Greenfield     & 8K                  & 2.4                      & 0.891                 & 168                        \\
Greenfield     & 8K                  & 2.0                      & 0.866                 & 130                        \\
Greenfield     & 8K                  & 1.6                      & 0.826                 & 91                         \\
Greenfield     & 8K                  & 1.2                      & 0.79                  & 58                         \\ \bottomrule
\end{tabular}
\end{table*}

\begin{figure}[h]
    \centering
    \includegraphics[width=1.0\textwidth]{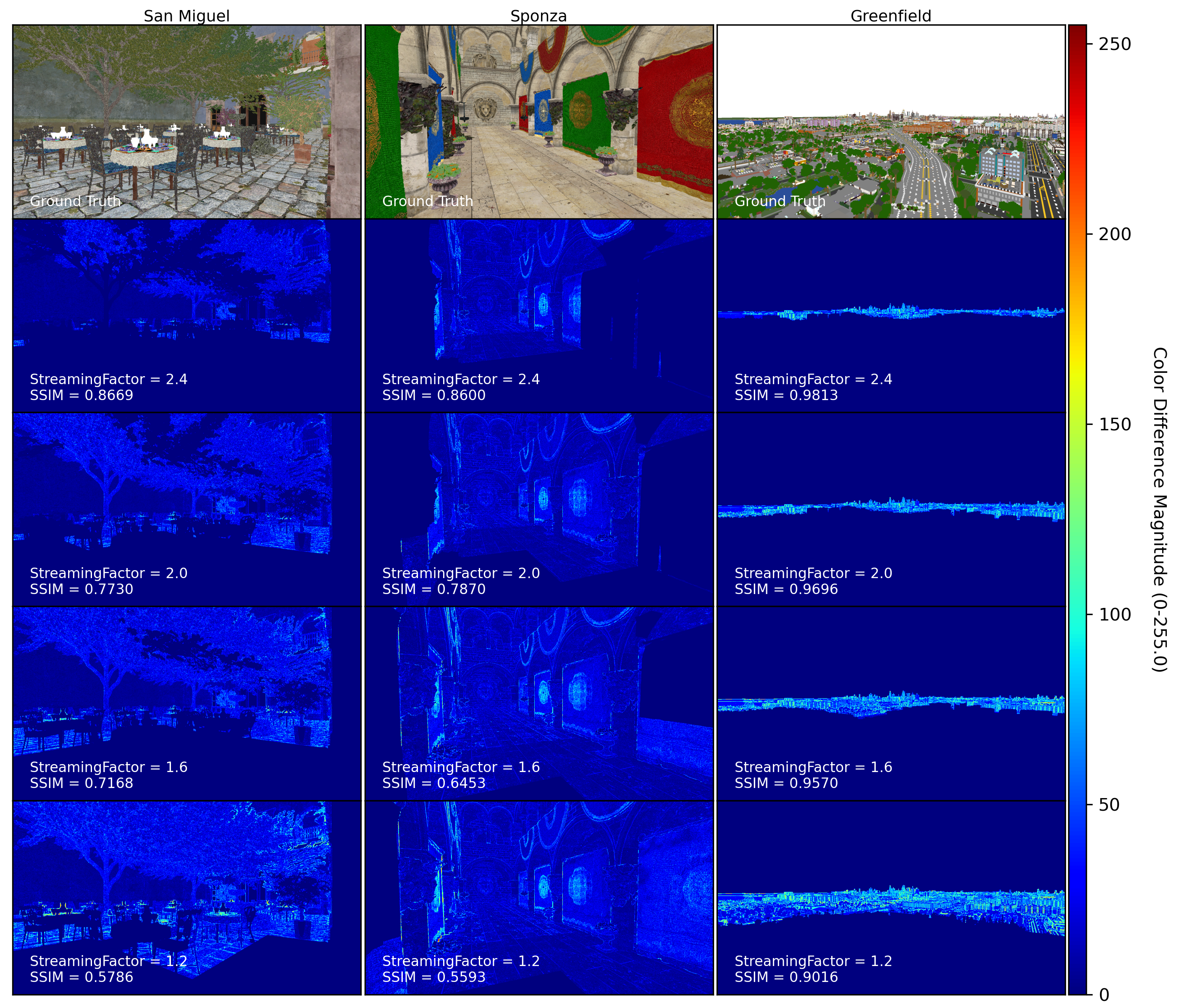}
    \caption{For each scene, we compare the rendering quality against the ground truth at different $StreamingFactor$ values. A higher $StreamingFactor$ results in finer LODs for distant regions and improved rendering quality.}
    \Description{For each scene, we compare the rendering quality against the ground truth at different $StreamingFactor$ values. A higher $StreamingFactor$ results in finer LODs for distant regions and improved rendering quality.}
    \label{fig:rendering_quality_cmp}
\end{figure}

\begin{figure}[h]
    \centering
    \includegraphics[width=0.9\textwidth]{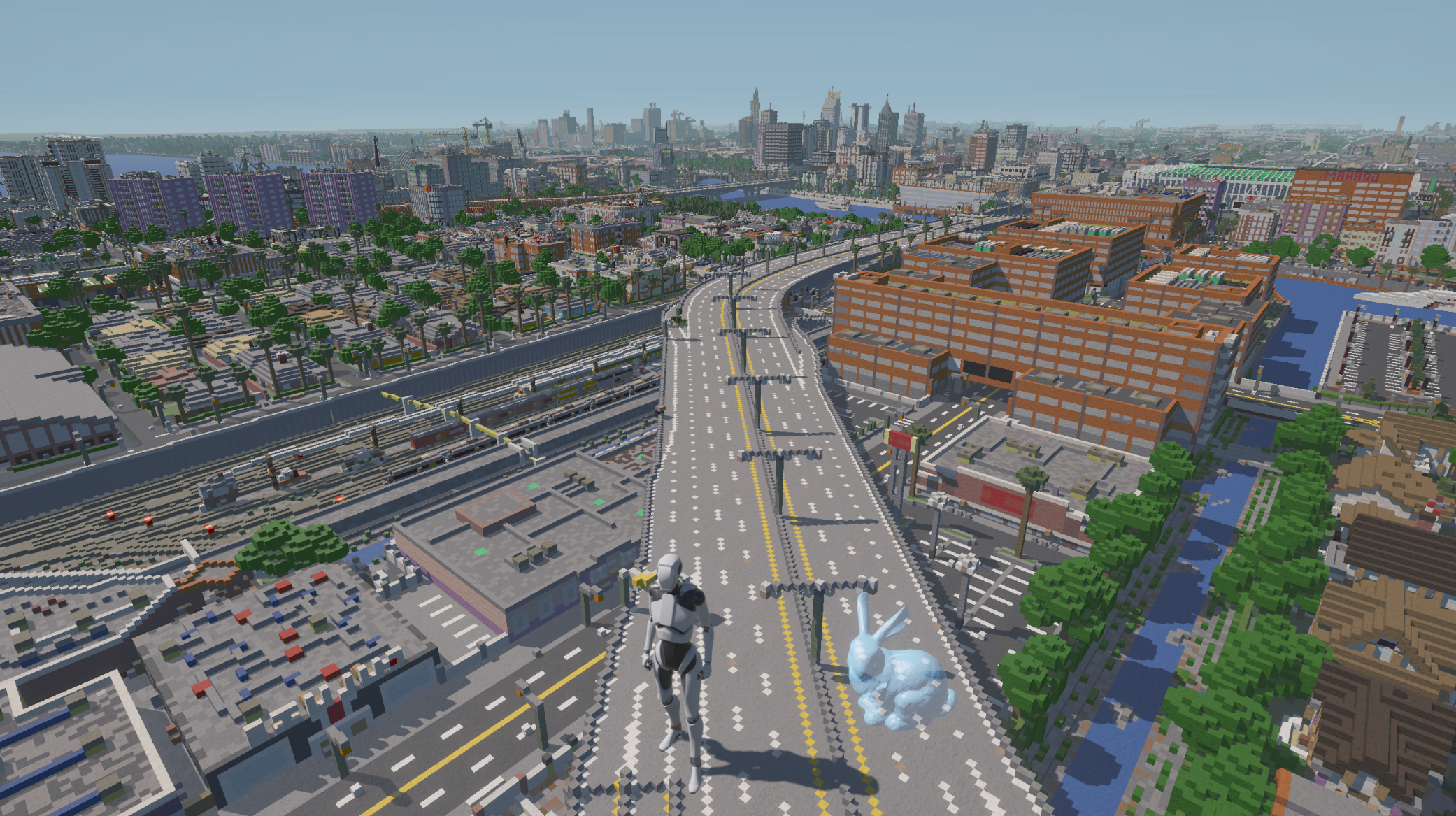}
    \caption{Our method can seamlessly integrate with the rendering of meshes. In the scene shown in the image, the environment is voxel-based, while the transparent bunny and the robot in the foreground are meshes.}
    \Description{The display of simultaneously rendering voxels and meshes.}
    \label{fig:game_scene_example}
\end{figure}

We visualized the differences between the rendering results of the scene under different StreamingFactor values and the Ground Truth in Figure \ref{fig:rendering_quality_cmp}. It can be observed that the geometric differences of the scene are small; however, for distant chunks, there are some color discrepancies compared to the ground truth. This is because, when constructing the LOD, we simply averaged the colors of the 8 voxels at the current LOD level to determine the color of a single voxel at the next LOD level, and we specified that the colors of the six faces of the voxel are the same. 
Though this approach can introduce a potential risk of degenerated color quality when there is rare significant color differences across the faces, the final rendering quality is almost always acceptable in voxel games, as demonstrated in Figure \ref{fig:game_scene_example}.

We also compared the visual effects of our method with the Distant Horizon MOD of Minecraft and the voxel editing software Avoyd (see Figure \ref{fig:lod_cmp}). We used the largest city in Minecraft, Greenfield, as the test scene, and the results indicate that our method produces the most detailed distant scenery.

\subsection{Discussions}

Since our method only inserts several passes between the opaque pass and the transparent pass of forward rendering, it seamlessly integrates with the rendering of meshes. 
This is crucial for actual game development, as not all objects in a voxel game are necessarily represented in voxel form. 
Our method allows game developers the flexibility to choose between rendering with voxels or meshes. 
Although our method does not support transparent voxels, in open-world scenes, most objects are opaque, with only a few, such as rivers and windows, being transparent. For these transparent objects, developers can opt to render transparent meshes as a substitute.

Currently, our implementation does not support runtime modification of voxels. 
We believe that whether to support runtime voxel modification and what solution to use for this should depend on the specific needs of game developers. 
If developers do not want players to modify voxels during gameplay, they can choose not to support runtime modifications and simply pre-build the game map through preprocessing. 
In games similar to Minecraft, where voxel resolution is low and players can only interact with voxels within a certain range, dense array can be used to maintain the area near the player, with meshes representing LOD 0 chunks and SVDAG representing chunks of LOD 1 and above. This allows for quick modifications of voxels by modifying meshes at runtime. When players leave these modified areas, the system can utilize idle time to reconstruct the SVDAG chunks of LOD 1 and above from the dense array.
For games that require support for higher-resolution voxels, using dense arrays for modifications may lead to memory bottlenecks. 
In such cases, developers can refer to the implementation in HashDAG, using persistent data structures to support interactive modifications of larger-scale voxels.

\section{Conclusions and Future Work}

In this paper, we proposed Aokana for real-time rendering of large-scale voxels in open-world voxel games. We implemented it in the Unity game engine. 

We divide the world into several chunks, compress each chunk using SVDAG, and develop a streaming loading system that significantly reduces VRAM consumption. 
Instead of rendering a single deep SVDAG, we use multiple shallow SVDAGs, which reduces memory access performance losses caused by non-contiguous jumps when traversing the SVDAG. 
Additionally, we developed a GPU-driven voxel rendering pipeline that utilizes Hi-Z occlusion culling and a visibility buffer to decrease overdraw and enhance rendering performance.

In our future work, we plan to support voxel rendering with various material models. 
We also plan to better integrate Aokana with the traditional mesh-based methods, to seamlessly support mesh-based collision detection, navigation systems and other modules.
A promising direction for voxel-based collision detection is to employ ray intersection in a compute shader.

\begin{acks}
We would like to express our sincere gratitude to the Greenfield Builder Team for their creation of the Greenfield Minecraft map, which facilitated various tests during our development. 
\end{acks}

\bibliographystyle{ACM-Reference-Format}
\bibliography{sample-base}
\end{document}